\documentclass[10pt,journal,compsoc]{IEEEtran}

\usepackage{subfigure}
\usepackage{graphicx}
\usepackage{multirow}
\usepackage{algorithm}
\usepackage{algorithmic}
\usepackage{url}
\usepackage{balance}
\usepackage{booktabs}
\usepackage{proof}
\usepackage{amsmath,amssymb,amsfonts}

\usepackage{threeparttable}

\newtheorem{definition}{Definition}

\ifCLASSOPTIONcompsoc
  \usepackage[nocompress]{cite}
\else
  \usepackage{cite}
\fi
\ifCLASSINFOpdf
\else
\fi
\begin{document}
%
\title{Deep Stable Multi-Interest Learning for Out-of-distribution Sequential Recommendation}
%
%
%
%

\author{Qiang~Liu,~\IEEEmembership{Member,~IEEE},
        Zhaocheng~Liu,
        Zhenxi~Zhu,
        Shu~Wu,~\IEEEmembership{Senior Member,~IEEE},
        and~Liang~Wang,~\IEEEmembership{Fellow,~IEEE}
\IEEEcompsocitemizethanks{
\IEEEcompsocthanksitem Qiang Liu and Zhaocheng Liu contribute equally to this work.
\IEEEcompsocthanksitem Qiang Liu, Shu Wu and Liang Wang are with the Center for Research on Intelligent Perception and Computing (CRIPAC), State Key Laboratory of Multimodal Artificial Intelligence Systems (MAIS), Institute of Automation, Chinese Academy of Sciences (CASIA), Beijing, China.\protect\\
E-mail: \{{qiang.liu, shu.wu, wangliang}\}@nlpr.ia.ac.cn
\IEEEcompsocthanksitem Zhaocheng Liu is with Kuaishou Technology, Beijing, China.\protect\\
E-mail: lio.h.zen@gmail.com
\IEEEcompsocthanksitem Zhenxi Zhu is with the Department of Computer Science and Technology, Nanjing University, Nanjing, China.\protect\\
E-mail: zhuzhenxi@smail.nju.edu.cn
}
\thanks{
(Corresponding author: Shu Wu)}}

%
%

\markboth{Journal of \LaTeX\ Class Files,~Vol.~14, No.~8, August~2015}%
{Shell \MakeLowercase{\textit{et al.}}: Bare Demo of IEEEtran.cls for Computer Society Journals}
%



\IEEEtitleabstractindextext{%
\begin{abstract}
Recently, multi-interest models, which extract interests of a user as multiple representation vectors, have shown promising performances for sequential recommendation.
However, none of existing multi-interest recommendation models consider the Out-Of-Distribution (OOD) generalization problem, in which interest distribution may change.
Considering multiple interests of a user are usually highly correlated, the model has chance to learn spurious correlations between noisy interests and target items.
Once the data distribution changes, the correlations among interests may also change, and the spurious correlations will mislead the model to make wrong predictions.
To tackle with above OOD generalization problem, we propose a novel multi-interest network, named DEep Stable Multi-Interest Learning (DESMIL), which attempts to de-correlate the extracted interests in the model, and thus spurious correlations can be eliminated.
DESMIL applies an attentive module to extract multiple interests, and then selects the most important one for making final predictions.
Meanwhile, DESMIL incorporates a weighted correlation estimation loss based on Hilbert-Schmidt Independence Criterion (HSIC), with which training samples are weighted, to minimize the correlations among extracted interests.
Extensive experiments have been conducted under both OOD and random settings, and up to $36.8\%$ and $21.7\%$ relative improvements are achieved respectively.
\end{abstract}

\begin{IEEEkeywords}
Sequential recommendation, multi-interest, out-of-distribution, stable learning.
\end{IEEEkeywords}}

\maketitle

\IEEEdisplaynontitleabstractindextext

%
\IEEEpeerreviewmaketitle

\section{Introduction}
\label{section:intro}
\IEEEPARstart{S}{equential} recommender systems aim to predict the next item(s) that a user might be interested in based on historical interactions.
It has become a vital research topic of recommender systems, in scenarios such as online shopping, online video and restaurant visiting.
Given historical behaviors, accurately capturing users' dynamic preferences is the core concern of sequential recommendation.
Nowadays, plenty of solutions based on recurrent neural networks \cite{yu2016dynamic,hidasi2015session}, convolutional neural networks \cite{tang2018personalized} and attentive networks \cite{liu2018stamp,kang2018self,sun2019bert4rec} have been proposed.

\begin{figure*}
	\centering
	\subfigure[Training on data collected from \emph{basketball and football} fans group, and inference on data collected from \emph{basketball and tennis} fans group.]{
		\begin{minipage}[b]{0.48\textwidth}
			\includegraphics[width=1\textwidth]{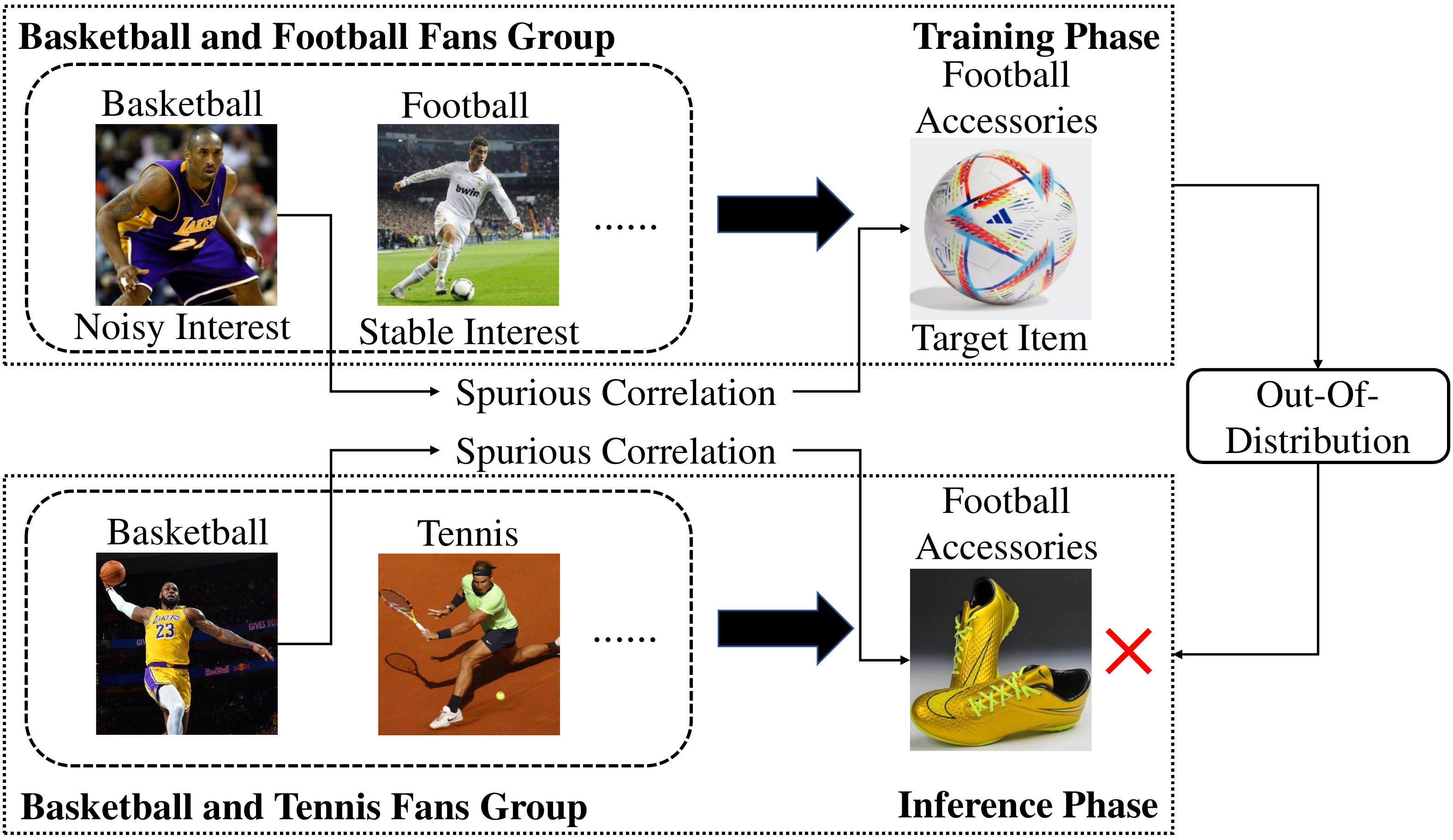}
		\end{minipage}
	\label{fig:examples1}
	}
	\subfigure[Training on data collected during World Cup, and inference on data collected after World Cup.]{
		\begin{minipage}[b]{0.48\textwidth}
			\includegraphics[width=1\textwidth]{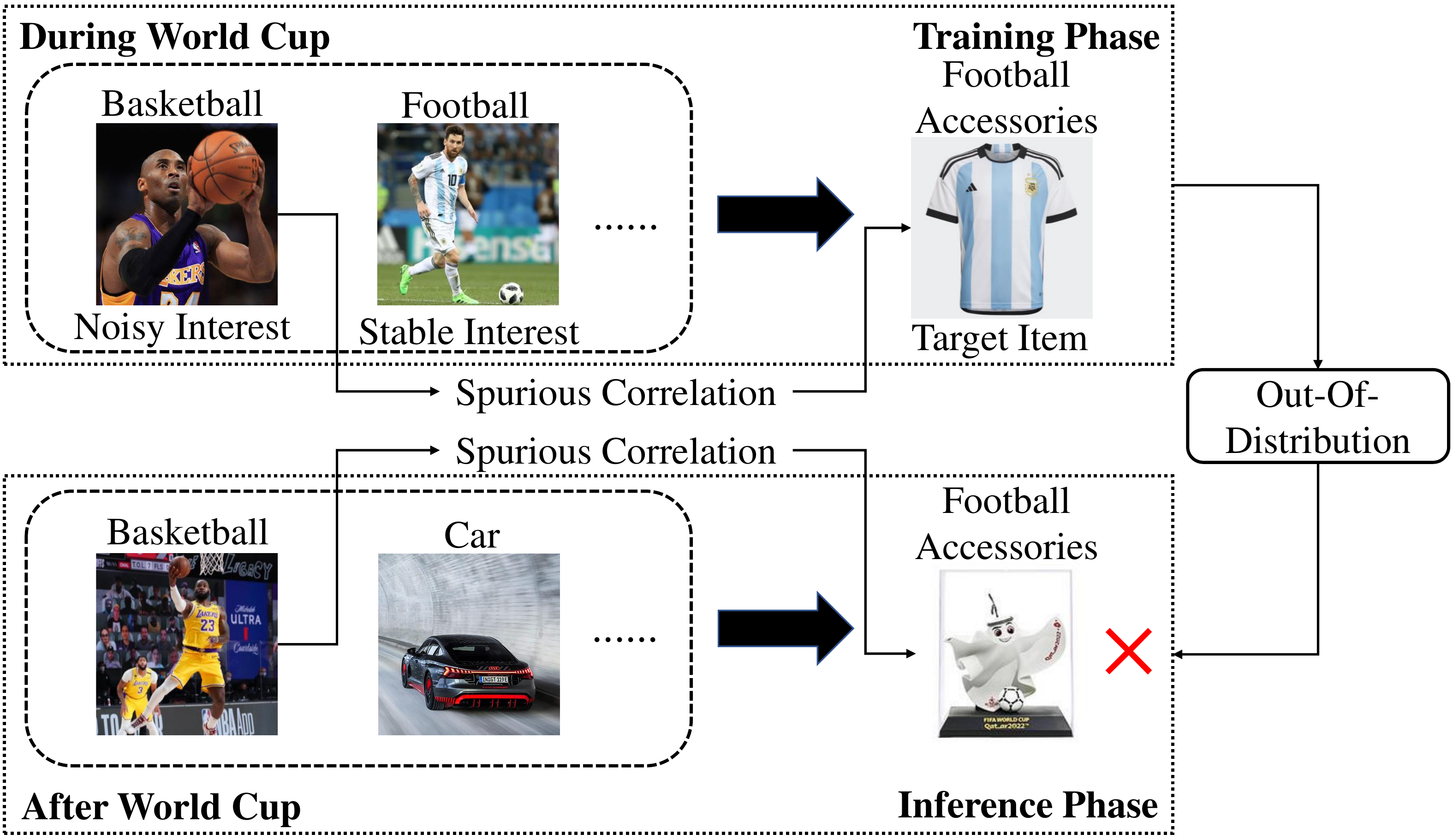}
		\end{minipage}
	\label{fig:examples2}
	}
	\caption{We show two examples for demonstrating the OOD generalization problem in multi-interest recommendation models. Conventional models may mistakenly learn spurious correlations between noisy interests and target items (\emph{basketball} and \emph{football accessories} in the figure) during training phase, and thus mislead the model predictions during inference phase.}
	\label{fig:examples}
\end{figure*}

For accurate sequential recommendation, we need to look into users' behavior history, and capture their interests on different types of items.
Meanwhile, a user usually has multiple interests.
For example, on an online video platform, a user may like to watch football videos, funny videos and car videos at the same time.
For modeling users' multiple interests, multi-interest networks \cite{li2019multi,cen2020controllable,tan2021sparse,tian2022multi} have been proposed to encode multiple interests with multiple representation vectors.
However, existing multi-interest approaches ignore to consider that the interest distribution in a system is always changing.
This brings multi-interest models the Out-Of-Distribution (OOD) problems \cite{shen2021towards,zhang2021causerec}, which may inevitably harm the performances of multi-interest recommenders when data distribution is changing.
To make the OOD generalization problem in multi-interest models clear, we present Def. \ref{def:stable_interests}-\ref{def:stable_model}.

\begin{definition}(Stable Interests) \label{def:stable_interests}
Given a target item and a user's behavior history, stable interests are those have causal relations to the target item.
We denote stable interests as $\left( {\mathop s\nolimits_0 ,\mathop s\nolimits_1 ,\mathop s\nolimits_2 ,...} \right)$, and target item as $y$.
Then, the above process can be denoted as $\left( {\mathop s\nolimits_0 ,\mathop s\nolimits_1 ,\mathop s\nolimits_2 ,...} \right) \to y$.
\end{definition}

\begin{definition}(Noisy Interests) \label{def:noisy_interests}
Given a target item and a user's behavior history, noisy interests are those have no causal relations to the target item.
We denote noisy interests as $\left( {\mathop n\nolimits_0 ,\mathop n\nolimits_1 ,\mathop n\nolimits_2 ,...} \right)$.
Then, the above process can be denoted as $\left( {\mathop n\nolimits_0 ,\mathop n\nolimits_1 ,\mathop n\nolimits_2 ,...} \right) \perp y$.
And there are usually unstable dependencies between stable interests and noisy interests.
, which can be denoted as $\left( {\mathop s\nolimits_0 ,\mathop s\nolimits_1 ,\mathop s\nolimits_2 ,...} \right) \to \left( {\mathop n\nolimits_0 ,\mathop n\nolimits_1 ,\mathop n\nolimits_2 ,...} \right)$.
\end{definition}

\begin{definition}(Unstable Multi-interest Model) \label{def:unstable_model}
Given a limited training set $\mathop \Omega \nolimits_{train}$, for conducting prediction ${\hat y}$, an unstable multi-interest model has chance to learn both causal relation $\left( {\mathop s\nolimits_0 ,\mathop s\nolimits_1 ,\mathop s\nolimits_2 ,...} \right) \to \hat y$ and spurious correlation $\left( {\mathop n\nolimits_0 ,\mathop n\nolimits_1 ,\mathop n\nolimits_2 ,...} \right) \to \hat y$.
Unstable multi-interest models tend to fail in OOD environments.
Spurious correlations exist due to the dependencies between stable interests and noisy interests.
\end{definition}

\begin{definition}(Stable Multi-interest Model) \label{def:stable_model}
Given a limited training set $\mathop \Omega \nolimits_{train}$, for conducting prediction ${\hat y}$, a stable multi-interest model learns causal relation $\left( {\mathop s\nolimits_0 ,\mathop s\nolimits_1 ,\mathop s\nolimits_2 ,...} \right) \to \hat y$, and eliminates spurious correlations, i.e., $\left( {\mathop n\nolimits_0 ,\mathop n\nolimits_1 ,\mathop n\nolimits_2 ,...} \right) \perp \hat y$.
It is necessary for stable multi-interest models to generalize to different OOD environments.
\end{definition}

The shift of interest distribution may be caused by 
(1) the user distribution is constantly changing with the development of a platform, and new users keep appearing;
(2) in the same user group, the interest distribution is also changing, due to changing popularity trends or recommendation strategies.
As shown in Fig. \ref{fig:examples}, we demonstrate some examples.
The first example in Fig. \ref{fig:examples1} is about the distribution shift between two different user groups.
And the second example in Fig \ref{fig:examples2} is about the distribution shift between the time periods during and after World Cup, which is a popular trend that affects user behaviors.
Considering a user usually has relatively similar interests, multiple interests extracted from the user's behavior history tend to be correlated.
And online systems, which tend to recommend items similar to the user's historical interests, further aggravate the dependencies between stable interests and noisy interests.
With such dependencies, we have chance to learn not only causal relations between stable interests and target items, but also spurious correlations between noisy interests and target items.
In both examples in Fig. \ref{fig:examples}, the dependencies between \emph{football} and \emph{basketball} may affect the model training process, and make multi-interest recommendation models learn the spurious correlation between \emph{basketball} and \emph{football accessories}.
Dependencies between stable interests and noisy interests are unstable, and may change in different data distribution.
Once the data distribution changes, i.e., the dependency between \emph{football} and \emph{basketball} in the behavior history changes, the spurious correlation will make the model produce wrong predictions during inference phase.
Accordingly, we have to study stable multi-interest recommendation models which can generalize to different ODD environments.

According to above analysis and examples, to alleviate the OOD generalization problem in multi-interest recommendation models, we need to remove the dependencies between stable interests and noisy interests, which limits the model to learn the causal relations between stable interests and target items.
To do this, Inverse Propensity Weighting (IPW) approaches \cite{schnabel2016recommendations,wang2019doubly,arbour2021permutation,wang2022unbiased} can be adopted.
However, propensity scores are hard to accurately estimate, and the variance of estimation is usually high \cite{zhang2021causal}.
Meanwhile, some work attempt to generate counterfactual sample for training causal models \cite{zhang2021causerec,wang2021counterfactual}.
But the quality of counterfactual sample generation is difficult to guarantee, and such methods lack enough explore space.
Moreover, it is hard to accurately distinguish stable interests and noisy interests from the multiple interests in the model, which makes the dependencies between stable interests and noisy interests hard to remove.
To tackle with above difficulties, we draw lessons from stable learning \cite{shen2020stable,kuang2018stable,kuang2020stable,kuang2021balance}.
Instead of identifying stable interests and noisy interests, we can remove the correlations among all the interests in the model.
To achieve this, we can incorporate a correlation estimation loss, which can be optimized together with the main objective.

Formally, in this paper, we propose a novel multi-interest network, named DEep Stable Multi-Interest Learning (DESMIL).
(1) DESMIL constructs a multi-interest extractor based on attention \cite{vaswani2017attention}, and use it to extract multi-interest representations from input user behavior sequence.
(2) Then, DESMIL selects the most important interest from the extracted ones as the representation of a user, and use it for making final predictions and constructing the main objective loss for model training.
(3) Meanwhile, DESMIL incorporates a weighted correlation estimation loss. To estimate degree of correlations among the extracted interest, we adopt Hilbert-Schmidt Independence Criterion (HSIC) \cite{gretton2005measuring,gretton2007kernel} which is a widely-used non-linear independent testing statistic and has been applied for feature de-correlation \cite{bahng2020learning}. We assign a weight for each sample, and obtain weighted HSIC as the weighted correlation estimation loss. DESMIL minimizes the weighted correlation estimation loss via optimizing the sample weights.
(4) Then, the sample weights are also added to the main objective loss, and a weighted main objective loss is obtained. That is to say, samples with higher degrees of correlations among interests tend to have lower weights for model training, and vice versa. This makes the multi-interest model actually being trained on a weighted training dataset, in which correlations among different interests are minimized \cite{zhang2021deep,fan2021generalizing}.
(5) Finally, the two losses, i.e., the weighted main objective loss and the weighted correlation estimation loss, are optimized iteratively, until convergence is reached.
To this end, it is able for the DESMIL model to better learn the causal relations between stable interests and target items.
We have conducted extensive experiments on three real-world datasets under both OOD and random experimental settings, in which DESMIL achieves promising results.

To summarize, the main contributions of this paper are listed as follows:
\begin{itemize}
\item We for the first time analyze the OOD generalization problem in multi-interest recommendation models, and propose to de-correlate different interests. This enables the multi-interest recommendation model to learn the causal relations between stable interests and target items, and eliminate spurious correlations between noisy interests and target items.
\item We propose a novel DESMIL model, which learns stable representations for sequential recommendation, and makes stable and accurate predictions generalized to OOD environments.
\item Extensive experiments show that our proposed DESMIL model outperforms several state-of-the-art sequential recommendation models by a significant margin, especially under the OOD setting.
\end{itemize}

The rest of the paper is organized as follows.
In Section 2, we review some related work on sequential recommendation, deep multi-interest models and stable learning.
Then we analyze the causal view of multi-interest recommendation models, and introduce the statistical criterion for non-liner correlation estimation in Section 3.
Sections 4 details our proposed DESMIL model.
In Section 5, we conduct empirical experiments to verify the effectiveness of DESMIL.
Section 6 concludes our work.

\section{Related Work}

In this section, we review some works on sequential recommendation, deep multi-interest models, and stable learning.

\subsection{Sequential Recommendation}

Modeling users’ dynamic preferences from historical behaviors is the core concern of research in sequential
recommendation \cite{fang2020deep}, which is a major task in recommender systems.
In some traditional models \cite{rendle2010factorizing,he2016fusing,he2017translation,hidasi2016general}, Markov chain and matrix factorization are exploited to model historical behaviors.
The most representative model is FPMC \cite{rendle2010factorizing}, which adopts a personalized Markov chain and train the model with a factorization model for capturing collaborative information.
In recent years, various deep neural networks such as recurrent neural network \cite{yu2016dynamic,hidasi2015session,liu2016predicting,liu2016context,liu2017multi}, convolutional neural networks \cite{tang2018personalized,wang2019towards} and attention-based networks \cite{liu2018stamp,kang2018self,sun2019bert4rec,luo2021stan,li2020time,hsu2021retagnn,fan2022sequential} have been exploited in deep sequential recommendation models.
Target-aware attention for conduction recommendation has also been studied \cite{zhou2018deep,zhou2019deep}.
Recently, contrastive learning has been applied in sequential recommendation \cite{zhou2020s3,xie2020contrastive,liu2021contrastive,chen2022intent}, for dealing with sparsity and noise in data.
Meanwhile, some works \cite{cui2018mv,liu2021noninvasive,xie2022decoupled} attempt to leverage variety side information for sequential recommendation.

Meanwhile, causal inference have been investigated for increasing the causality and eliminate biases in recommendation \cite{schnabel2016recommendations,wang2019doubly,wang2022unbiased,zhang2021causal,wang2021counterfactual,mu2022alleviating,wang2022causal,wang2022escm2,wang2021clicks}.
Among them, biases such as exposure bias \cite{schnabel2016recommendations,wang2019doubly,guo2021enhanced,wang2022escm2} and popularity bias \cite{zhang2021causal,wei2021model} are widely studied in static recommenders.
The user demographic feature shift problem has also been studied with causal representation learning for collaborative filtering \cite{wang2022causal}.
For sequential recommendation, the exposure bias is also studied to deal with the missing-not-at-random problem in the user behavior history \cite{wang2022unbiased,damak2022debiasing,xu2022dually}. 
And some works \cite{zhang2021causerec,wang2021counterfactual,yang2021top} generate counterfactual samples in sequential user behavior history, for training models generalizable to OOD environments.

\subsection{Deep Multi-interest Models}

In real scenarios, a user may have multiple interests in the behavior history, and an overall user preference representation as in most models can hardly grasp the diverse essence of user interests \cite{liu2019single,tan2021sparse}.
So, we need to extract multiple interests of a user from the behavior history for better sequential recommendation.
There are some work \cite{li2019multi,cen2020controllable,tan2021sparse,tian2022multi,chen2020improving,ma2019learning} studying how to effectively extract a user's multiple interests in sequential recommendation as multiple vectors.
MIND \cite{li2019multi} firstly proposes a multi-interest extractor based on the dynamic routing mechanism \cite{sabour2017dynamic,hinton2018matrix,hinton2011transforming}. 
As the procedure of dynamic routing can be seen as soft-clustering, the user's historical behaviors can be grouped into different clusters. 
Meanwhile, a label-aware attention mechanism is proposed to effectively aggregate the multiple user preference representations in training.
Besides, ComiRec \cite{cen2020controllable} proposes a controllable multi-interest Framework, in which, both dynamic routing and attentive models can be adopted to extract multiple user interests. 
Lately, instead of implicitly generating a user's multiple interests by clustering the user behaviors, SINE \cite{tan2021sparse} directly maintains a pool of conceptual prototypes to represent the all set of the user's potential interests. 
Then a self-attention mechanism is used to decide which prototypes are activated to the user's multiple interests.
MGNM \cite{tian2022multi} proposes to combine multi-interest learning modal and graph convolutional networks.
By aggregating multi-level user preferences, MGNM extracts a user's multiple interests more precisely.
Meanwhile user-aware candidate matching is also studied in multi-interest models \cite{chai2022user}.

\subsection{Stable Learning}

The out-of-distribution problem \cite{shen2021towards} is a common challenge in real-world scenarios, and stable learning has become a successful way to deal with this recently.
Stable learning aims to learn a stable predictive model that achieves uniformly good performance on any unknown test data \cite{kuang2018stable}.
The framework of most stable learning works can be divided into two steps: sample weight learning and weighted training.
Specifically, sample weights are learned to de-correlate features in training data, and then weighted training is conducted to train models on weighted feature distribution, which is an approach to independent identically feature distribution.
Along this strand, various de-correlation methods \cite{shen2020stable,kuang2018stable,kuang2020stable,kuang2021balance} have been proposed to learn sample weights and train linear stable models.
Moreover, StableNet \cite{zhang2021deep} proposes to adopt random Fourier features to eliminate non-linear dependencies among features in convolutional neural networks.
And feature de-correlation in graph neural networks \cite{fan2021generalizing,fan2022debiased,li2022ood} and healthcare \cite{luo2022deep} has also been studied.
Lately, Xu et al.\cite{xu2021stable} theoretically proves that the stability of least square regression and binary classification can be guaranteed with mutual independence of feature variables under mild conditions.

\section{Preliminaries} \label{section:preliminaries}

In this section, we analyze the causal view of multi-interest recommendation models, and introduce the statistical criterion for non-liner correlation estimation.

\begin{figure}
	\centering
	\subfigure[Ideal causal diagram.]{
		\begin{minipage}[b]{0.14\textwidth}
			\includegraphics[width=1\textwidth]{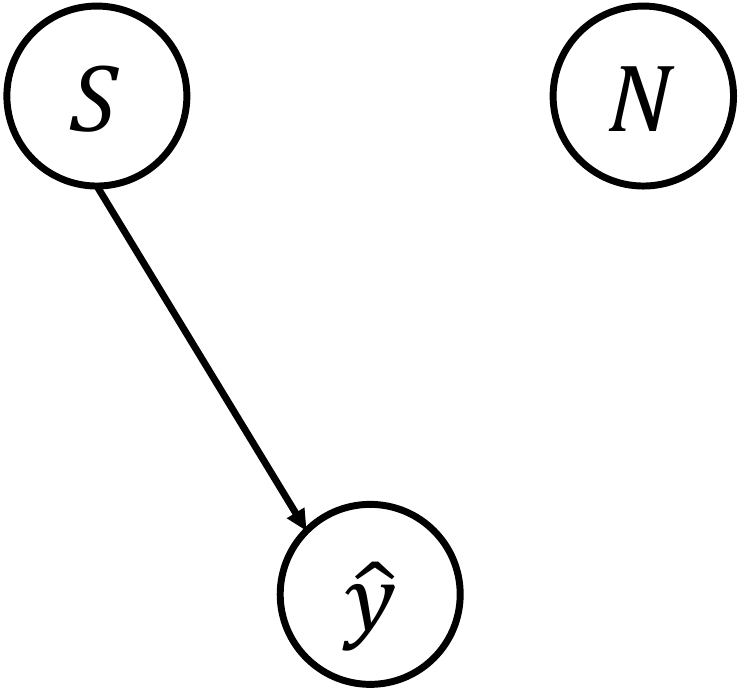}
		\end{minipage}
	\label{fig:causal1}
	}
	\subfigure[Causal diagram of conventional models.]{
		\begin{minipage}[b]{0.14\textwidth}
			\includegraphics[width=1\textwidth]{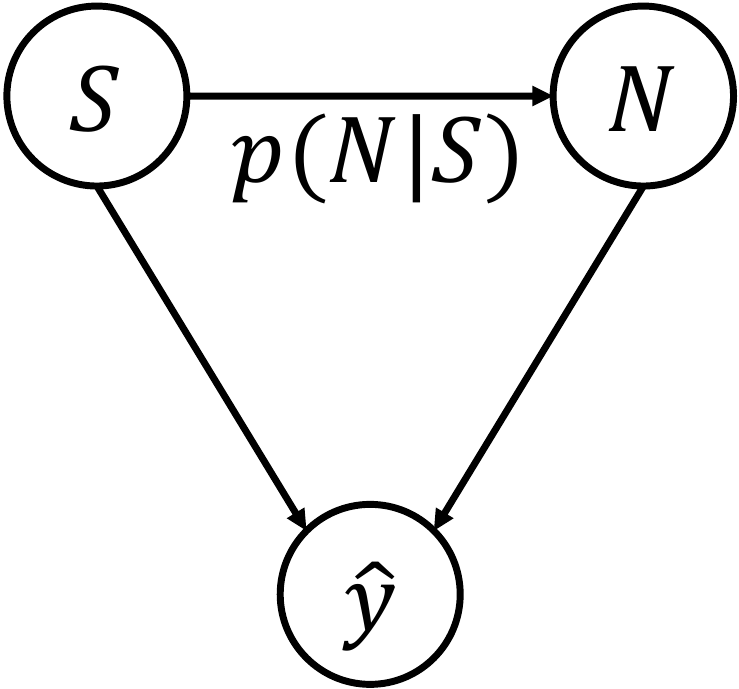}
		\end{minipage}
	\label{fig:causal2}
	}
	\subfigure[Removing dependency $S \to N$.]{
		\begin{minipage}[b]{0.14\textwidth}
			\includegraphics[width=1\textwidth]{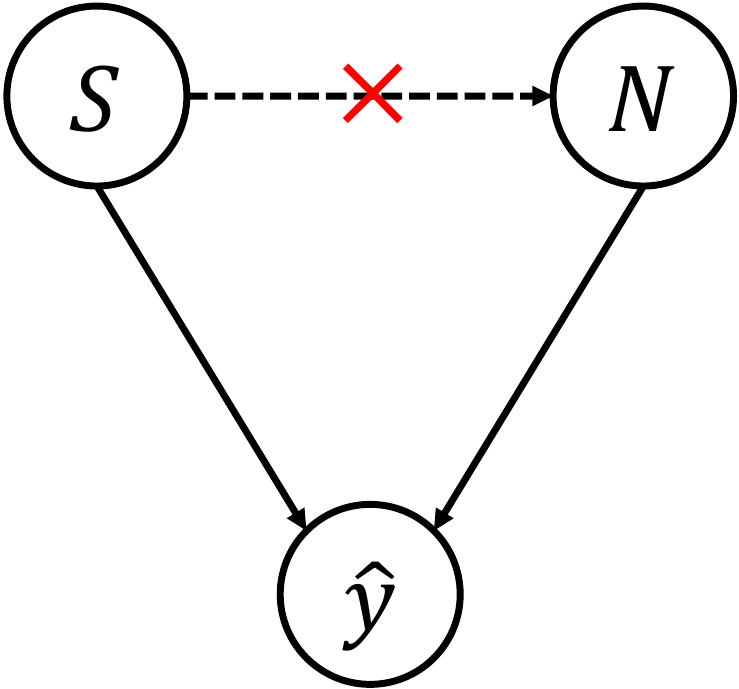}
		\end{minipage}
	\label{fig:causal3}
	}
	\caption{Causal view analysis of multi-interest recommendation models.}
	\label{fig:causal}
\end{figure}

\subsection{Causal View of Multi-interest Models} \label{section:analysis}

In Fig. \ref{fig:causal}, we illustrate the causal diagrams of multi-interest sequential recommendation.
$S = \left( {\mathop s\nolimits_0 ,\mathop s\nolimits_1 ,\mathop s\nolimits_2 ,...} \right)$, $N = \left( {\mathop n\nolimits_0 ,\mathop n\nolimits_1 ,\mathop n\nolimits_2 ,...} \right)$ and $\hat y$ denote stable interests, noisy interests and prediction of target item respectively.
Fig. \ref{fig:causal1} is the ideal causal diagram, in which only the causal relation $S \to \hat y$ exists.
However, as shown in Fig. \ref{fig:causal2}, dependency $S \to N$ exists, and leads conventional multi-interest recommendation models to learn spurious correlation $N \to \hat y$.
The dependency $S \to N$ is not stable, and $p\left( {N|S} \right)$ may vary in different environments.
Once the data distribution changes across training phase and inference phase (validating or testing), there will be $\mathop p\nolimits_{{\rm{train}}} \left( {N|S} \right) \ne \mathop p\nolimits_{{\rm{infer}}} \left( {N|S} \right)$, and the spurious correlation $N \to \hat y$ will mislead the model to make wrong predictions.
Examples for demonstrating above phenomenon can be found in Fig. \ref{fig:examples}.
Therefore, the path $N \leftarrow S \to \hat y$ establishes spurious correlation $N \to \hat y$.
Thus, as shown in Fig. \ref{fig:causal3}, we need to remove $S \to N$, so that spurious correlation $N \to \hat y$ can be eliminated, and causal relation $S \to \hat y$ can be accurately learned.

However, it is hard to accurately distinguish stable interests and noisy interests from the multiple interests in the model.
So, it is hard to directly perform interest de-correlation between $S$ and $N$.
Inspired by stable learning \cite{shen2020stable,kuang2018stable,kuang2020stable,kuang2021balance}, which de-correlates all the input features via sample re-weighting, we can remove the correlations among all the interests in the recommendation model instead.
That is to say, we can estimate the degree of correlation between each pair of interests, and minimize the degree of overall correlations.
Moreover, considering the correlation between two interest representation vectors shall be non-linear, we need a non-linear correlation estimation criterion.

\begin{figure}
    \centering
    \includegraphics[width=0.8\linewidth]{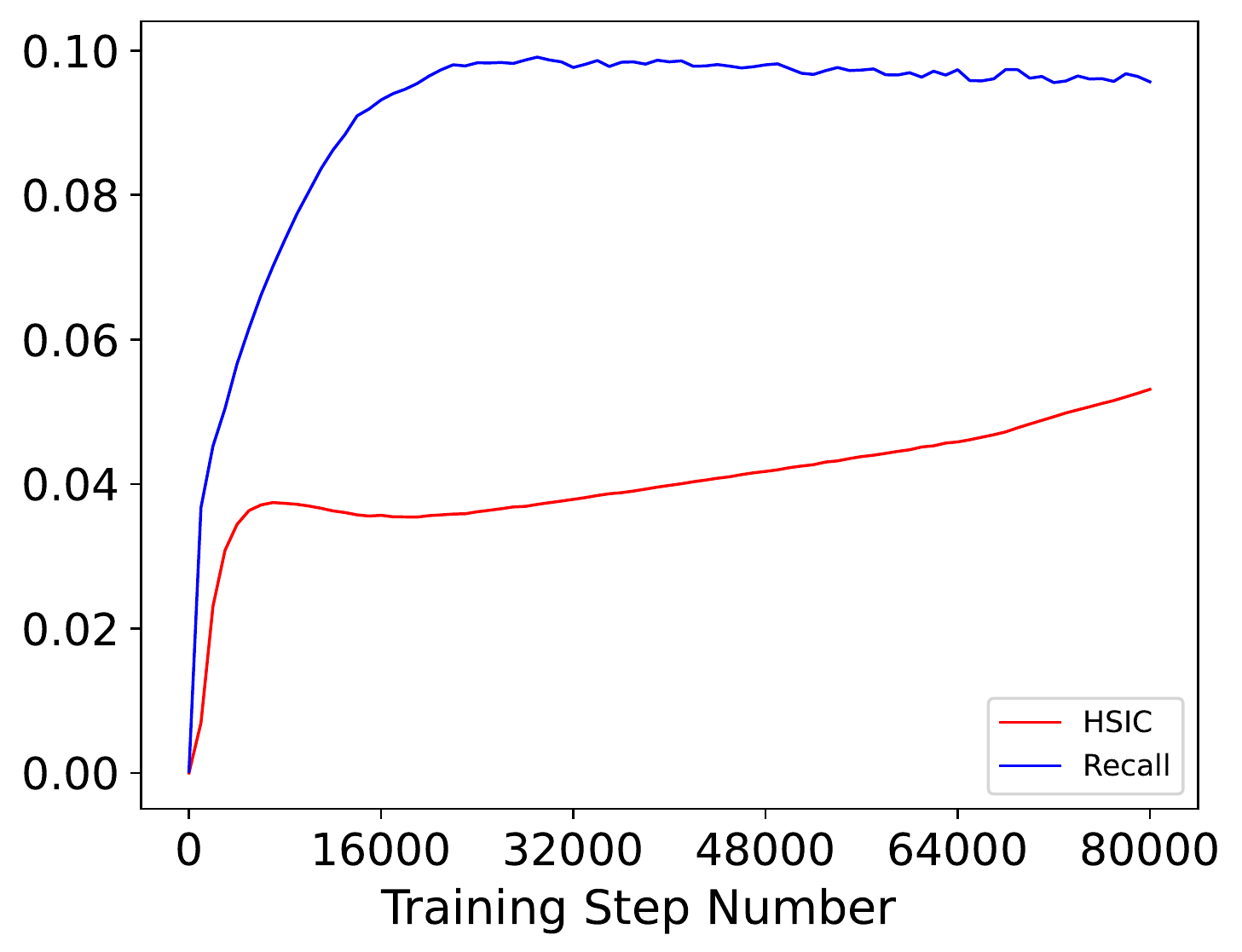}
    \caption{The curves of HSIC value on the training set and Recall@50 value on the validation set during training ComiRec on the Book dataset. From the latter parts of the curves, we can conclude that, the excessive correlations among interests limit the performance to further increase.}
    \label{fig:book_pilot}
\end{figure}

\begin{figure*}
    \centering
    \includegraphics[width=0.9\linewidth]{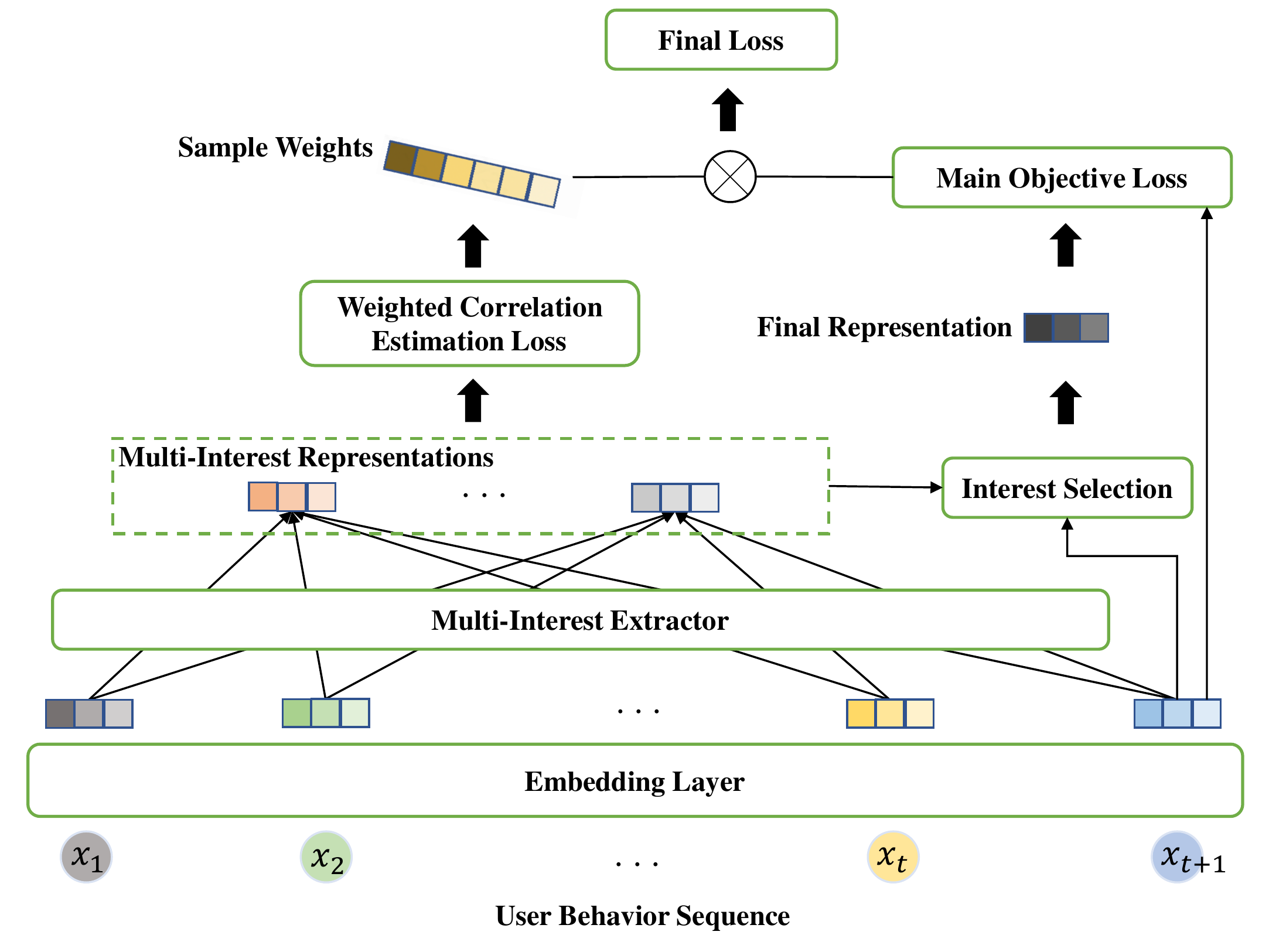}
    \caption{The overview of our proposed DESMIL model. DESMIL extracts multi-interest representations, and selects the most important one for making final predictions and constructing the main objective loss. Meanwhile, DESMIL incorporates a weighted correlation estimation loss based on HSIC, with which training samples are weighted, to minimize the correlations among extracted interests.}
    \label{fig:overview}
\end{figure*}

\subsection{Correlation Estimation Criterion} \label{section:criterion}

As we need to minimize correlations among multiple interests in the model, we have to estimate the degree of non-linear correlations between each pair of interests.
To achieve this, we adopt HSIC \cite{gretton2005measuring,gretton2007kernel}, which can estimate non-linear correlation between two variables and has been applied for feature de-correlation \cite{bahng2020learning}, as our correlation estimation criterion.
HSIC is the Hilbert-Schmidt norm of the cross-covariance operator between the distributions in Reproducing Kernel Hilbert Space (RKHS).

Given two random variables $U$ and $V$, the formulation of HSIC is:
\begin{equation}
    \begin{aligned}
    HSIC(U, V)& = \mathbb{E}_{uu'vv'}[k_u(u, u')k_v(v, v')] \\
    &+ \mathbb{E}_{uu'}[k_u(u, u')]\mathbb{E}_{vv'}[k_v(v,v')] \\
    &- 2\mathbb{E}_{uv}[\mathbb{E}_{u'}(k_u(u, u'))\mathbb{E}_{v'}[k_v(v,v')]],
    \end{aligned}
\end{equation}
where $\mathbb{E}_{uu'vv'}$ denotes the expectation over independent pairs $(u, v)$ and $(u', v')$ drawn from $P(U, V)$, $k_u$ and $k_v$ are kernel functions.
We use the Radial Basis Function (RBF) kernel which is formulated as:
\begin{equation}
    k(u,v) = exp\left(-\frac{||u-v||^2_2}{\sigma^2}\right).
\end{equation}
Given $m$ samples drawn from $P(U, V)$, the Empirical HSIC \cite{gretton2005measuring} is defined as
\begin{equation}
    HSIC(U,V) = (m-1)^{-2}tr(\mathbf{K}_U\mathbf{P}\mathbf{K}_V\mathbf{P}),
\end{equation}
where $\mathbf{K}_U \in \mathbb{R}^{m\times m}$ and $\mathbf{K}_V \in \mathbb{R}^{m\times m}$ have entries $\mathbf{K}_{U_{ij}} = k(U_i,U_j)$ and $\mathbf{K}_{V_{ij}} = k(V_i,V_j)$, $\mathbf{P} = \mathbf{I} - \frac{1}{m}\mathbf{1}\mathbf{1}^\mathbb{T} \in \mathbb{R}^{m\times m}$ is the centering matrix, and $\mathbf{1}$ is an $(m \times 1)$-dimensional vector of ones.
To be noted, $HSIC(U, V) = 0$ if and only if $U \perp V$.

Furthermore, to investigate the impact of correlations among extracted interests to the training process of multi-interest recommendation models, in Fig. \ref{fig:book_pilot}, 
we visualize the change of HSIC value on the training set and Recall@50 value on the validation set when training ComiRec \cite{cen2020controllable} on the Book dataset \cite{mcauley2015image,he2016ups}.
The HSIC value is calculated as the sum of non-linear correlations measured by HSIC between each pair of interest representations extracted in ComiRec.
We can observe from the curves that, after about 2000 steps, the value of HSIC keeps increasing slowly, while the value of recall@50 stops increasing and even begins to decrease.
That is to say, in the latter parts of the curves, the excessive correlations among interests limit the performance to further increase.
To some extent, for multi-interest models, above observation reveals the trade-off between the correlations among interests and the model performances.

\section{Methodology} \label{section:method}

In this section, we formulate the problem and introduce the proposed DESMIL model in detail, and the overview of DESMIL is illustrated in Fig. \ref{fig:overview}.

\subsection{Problem Formulation}

In the setting of sequential recommendation, we have a set of users $\mathcal{U} = \{u_1, u_2,\cdots, u_{|\mathcal{U}|}\}$ and a universe of items $\mathcal{I} = \{i_1, i_2,\cdots, i_{|\mathcal{I}|}\}$.
For each user $u$, given the behavior sequence $X^u = (x^u_1,x^u_2,\cdots,x^u_t)$ until time step $t$, we need to predict the user's next interaction $x^u_{t+1}$, where $x^u_t \in \mathcal{I}$.
The goal of sequential recommendation is to recommend to each user a list of items that maximize her/his future needs.
Meanwhile, in multi-interest models, there are multiple interest representations for each user, and we use $c$ to denote the number of interest representation vectors in the model.

\subsection{Multi-Interest Representation Extraction} \label{section:multi-interest}

Firstly, we embed items in the behavior sequence into dense vectors.
Specifically, given the input sequence $(x^u_1,\cdots,x^u_{t})$, we create an embedding matrix $\mathbf{V} \in \mathbb{R}^{\mathcal{|I|}\times d}$ where $d$ is the embedding dimensionality, and retrieve the input embedding matrix by applying the embedding look-up operation.
Besides, to make the proposed DESMIL aware of the positions of historical items, we inject the corresponding trainable position embedding matrix \cite{vaswani2017attention,kang2018self} $\mathbf{P} \in \mathbb{R}^{t\times d}$ into the input embedding matrix.
The final input embedding matrix $\mathbf{E}^u_t \in \mathbb{R}^{t\times d}$ can be formulated as
\begin{equation}
    \mathbf{E}^u_t = \left [ \begin{matrix}
    \mathbf{V}_{x^u_1} + \mathbf{P}_1 \\
    \vdots \\
    \mathbf{V}_{x^u_{t}} + \mathbf{P}_{t} \\
    \end{matrix} \right ].
\end{equation}

Then, we need a multi-interest extractor to generate multiple representation vectors to capture the diverse interests of a user.
Considering the specific interest extraction method is not our main concern, and the main objective of this work is to learn causal representations from multi-interest representations for OOD generalization, we empirically adopt a simple attentive module.
The attentive matrix $\textbf{A}^u_t \in \mathbb{R}^{c\times t}$ can be calculated as
\begin{equation}
    \mathbf{A}^u_t = \textnormal{softmax}(\mathbf{W}_2\tanh(\mathbf{W}_1(\mathbf{E}^u_t)^\top)),
\end{equation}
where $\mathbf{W}_1 \in \mathbb{R}^{\hat{d}\times d}$ and $\mathbf{W}_2 \in \mathbb{R}^{c\times\hat{d}}$ are trainable transformation matrices.
Then, we obtain the multi-interest representation matrix $\mathbf{M}^u_t \in \mathbb{R}^{c \times d}$ as 
\begin{equation}
    \mathbf{M}^u_t = \mathbf{A}^u_t \mathbf{E}^u_t.
\end{equation}
To this end, for each user, we obtain $c$ representation vectors to capture the diverse interests.

Furthermore, we adopt the interest selection strategy \cite{cen2020controllable} to choose the most important interest representation from captured interests to generate the final representation of the whole behavior sequence.
Given $x^u_{t+1}$ to be predicted, and its embedding $\mathbf{V}_{x^u_{t+1}}$, we generate the selected representation $\mathbf{R}^u_t \in \mathbb{R}^{1 \times d}$ as
\begin{equation}
    \mathbf{R}^u_t = \mathbf{M}^u_t[\mathop{argmax}(\mathbf{M}^u_t\mathbf{V}_{x^{u}_{t+1}}^{\top}),:].
\end{equation}

Finally, for a batch of samples $\mathcal{B}$ drawn from the training set $\mathop \Omega \nolimits_{train}$, the main objective loss is formulated as
\begin{equation} \label{eq:main_loss1}
    \mathcal{L}_{\rm{main}} = -\sum_{(u,t)\in \mathcal{B}}log\left(\frac{\mathop{exp}(\mathbf{R}^u_t\mathbf{V}_{x^{u}_{t+1}}^{\top})}{\sum_{i\in \mathcal{I}}\mathop{exp}(\mathbf{R}^u_t\mathbf{V}_i^\top)}\right),
\end{equation}
which can be implemented by the sampled softmax technique \cite{covington2016deep,jean2014using} considering computational efficiency.

\begin{algorithm}[t]
    \caption{Training process of DESMIL}
    \label{alg_training}
    \begin{algorithmic}[1]
        \REQUIRE Training set $\mathop \Omega \nolimits_{train}$, and maximum training epoch $\mathop{Epoch}$.
        \ENSURE Model parameters $\theta$.
        \STATE Initialize the iteration variable $q\gets 0$.
        \STATE Initialize the best iteration variable $q_{best}\gets 0$.
        \STATE Initialize sample weight $\omega^{(0)}_{u,t} = 1.0$, for all $u$ and $t$.
        \STATE Initialize model parameters $\theta^{(0)}$ via glorot uniform initializer \cite{glorot2010understanding}.
        \REPEAT
        \STATE Draw a batch of samples $\mathcal{B}$.
        \STATE $q\gets q+1$.
        \STATE Keep $\omega^{(q-1)}_{\mathcal{B}}$ fixed and minimize $\hat{\mathcal{L}}^{(q)}_{\rm{main}}$ via updating $\theta^{(q)}$, where $\hat{\mathcal{L}}^{(q)}_{\rm{main}}$ is defined in Eq. (\ref{eq:main_loss2}).
        \STATE  Keep $\theta^{(q)}$ fixed and update $\omega^{(q)}_{\mathcal{B}}$ via minimizing $\mathcal{L}^{(q)}_{\rm{corr}}$ as in Eq. (\ref{eq:de-correlation}), where $\mathcal{L}^{(q)}_{\rm{corr}}$ is defined in Eq. (\ref{eq:correlation_loss}).
        \STATE Update $q_{best}\gets q$, if better validation results achieved.
        \UNTIL{Early stopped or maximum training epoch is reached.}
        \RETURN $\theta^{(q_{best})}$.
    \end{algorithmic}
\end{algorithm}

\subsection{De-correlation among Multi-Interest Representations} \label{section: de-correlation}

As discussed in Sec. \ref{section:preliminaries}, we adopt sample re-weighting techniques \cite{kuang2020stable,shen2020stable,zhang2021deep} for de-correlation among multi-interest representations.
We propose an interest de-correlation regularizer that aims to estimate a weight for each sample, so that the degree of correlations among multiple interests can be minimized.
Specifically, we assign a weight $\omega_{u,t}$ for each sample.
We use $\omega^{(q)}_{u,t}$ to denote the sample weight after calculation of the $q$-th training epoch, and the initial sample weight as $\omega^{(0)}_{u,t} = 1.0$.

For de-correlation among multi-interest representations, we re-weight the main objective losses of samples in Eq. (\ref{eq:main_loss1}), and obtain the revised weighted main objective loss at epoch $q$ as
\begin{equation} \label{eq:main_loss2}
    \hat{\mathcal{L}}^{(q)}_{\rm{main}} = -\sum_{(u,t)\in \mathcal{B}}\omega^{(q-1)}_{u,t} log\left(\frac{\mathop{exp}(\mathbf{R}^u_t\mathbf{V}_{x^{u}_{t+1}}^{\top})}{\sum_{i\in \mathcal{I}}\mathop{exp}(\mathbf{R}^u_t\mathbf{V}_i^\top)}\right),
\end{equation}
which takes the samples weights optimized in the last epoch for re-weighting the model training process, and is illustrated as the final loss in Fig. \ref{fig:overview}.

Meanwhile, we need to estimate the sample weights via minimizing the degree of correlations among multiple interests.
Taking the weights from the last epoch, the multi-interest representation of user $u$ after time step $t$ is re-weighted as
\begin{equation}
    \hat{\mathbf{M}}^{(q)}_{u,t} = \omega^{(q-1)}_{u,t} \mathbf{M}^u_t.
\end{equation}
Then, we propose a weighted correlation estimation loss, which is based on HSIC introduced in Sec. \ref{section:criterion} and used to estimate the degree of correlations among multiple interests, as
\begin{equation} \label{eq:correlation_loss}
    \mathcal{L}^{(q)}_{\rm{corr}} = \sum_{(u,t)\in \mathcal{B}}\sum_{j}\sum_{k}\lambda HSIC(\hat{\mathbf{M}}^{(q)}_{u,t}[j,:], \hat{\mathbf{M}}^{(q)}_{u,t}[k,:]),
\end{equation}
where $\lambda$ is the de-correlation importance that controls the learning process of sample weights.
Then, via minimizing the degree of correlations, we optimize sample weights as
\begin{equation}\label{eq:de-correlation}
    \omega^{(q)}_{\mathcal{B}} = \mathop{argmin}\limits_{\omega}\mathcal{L}^{(q)}_{\rm{corr}}.
\end{equation}

Furthermore, we alternatively minimize the weighted main objective loss $\hat{\mathcal{L}}^{(q)}_{\rm{main}}$ with respect to sample weights $\omega^{(q)}$, and minimize the weighted correlation estimation loss $\mathcal{L}^{(q)}_{\rm{corr}}$ with respect to model parameters $\theta^{(q)}$.
Meanwhile, the detailed procedure of our proposed DESMIL model is shown in Alg. \ref{alg_training}.

Via the above process of training, samples with higher degrees of correlations among interests will have lower weights for model training in the main objective loss, and vice versa.
This makes the final model actually being trained on a weighted training dataset, in which correlations among different interests are minimized according to the loss in Eq. (\ref{eq:correlation_loss}).
To be noted, the sample weights are only optimized with samples in the training set, and can make the final model potentially generalize to OOD environments.
The inference procedure of DESMIL is the same as conventional multi-interest model, and we do not need to estimate sample weights for samples in the validation set or the testing set.

\begin{table*}[t]
\centering
\caption{Results under OOD data splitting evaluated by different metrics ($\%$). Best performances are indicated by bold font and the strongest baselines are underlined. The improvements indicate the relative increase of DESMIL over the best baselines.}
\label{tab:result_OOD}
    \begin{tabular}{cccccccccccc}
    \toprule
    Dataset & Metric & GRU4Rec & SASRec & MIND  & ComiRec & SINE  & MGNM  & USR  & CauseRec & DESMIL & Improv. \\
    \midrule
    \multirow{6}[2]{*}{Book} & Recall@20 & 3.24  & 4.95  & 5.36  & \underline{5.60} & 5.54  & 5.32  & 5.42  & 5.38  & \textbf{7.16} & 27.86\% \\
          & Recall@50 & 5.84  & 6.78  & 7.10  & \underline{7.96} & 7.56  & 7.35  & 7.65  & 7.71  & \textbf{10.89} & 36.81\% \\
          & NDCG@20 & 2.87  & 3.06  & 3.13  & 3.20  & 3.41  & 3.46  & \underline{3.55} & 3.25  & \textbf{4.25} & 19.72\% \\
          & NDCG@50 & 3.15  & 3.54  & 3.96  & 3.78  & 4.22 & 4.15  & \underline{4.30}  & 4.08  & \textbf{5.30} & 23.26\% \\
          & HR@20 & 6.93  & 8.25  & 9.98  & 10.25  & 10.17  & 10.26  & 10.31  & \underline{10.48} & \textbf{12.42} & 18.51\% \\
          & HR@50 & 12.10  & 13.56  & 15.98  & 15.84  & 15.86  & 15.74  & 15.48  & \underline{16.04} & \textbf{18.27} & 13.90\% \\
    \midrule
    \multirow{6}[2]{*}{Movies and TV} & Recall@20 & 10.87  & 12.34  & 13.14  & 13.65  & 13.64  & 13.79  & 13.46  & \underline{13.90} & \textbf{14.62} & 5.18\% \\
          & Recall@50 & 14.08  & 15.01  & 16.18  & 16.57  & 16.68  & 17.17  & 16.50  & \underline{17.33} & \textbf{18.65} & 7.62\% \\
          & NDCG@20 & 9.87  & 11.00  & 11.99  & 12.71  & 12.53  & 12.61  & 12.25  & \underline{12.91} & \textbf{13.87} & 7.44\% \\
          & NDCG@50 & 11.85  & 12.72  & 13.54  & 13.92  & 13.77  & 13.76  & 13.16  & \underline{14.40} & \textbf{15.44} & 7.22\% \\
          & HR@20 & 19.25  & 20.11  & 22.36  & 23.32  & 23.27  & 23.17  & 22.88  & \underline{23.44} & \textbf{23.77} & 1.41\% \\
          & HR@ 50 & 25.20  & 26.35  & 28.56  & 29.84  & 29.71  & 29.75  & 29.10  & \underline{30.05} & \textbf{31.00} & 3.16\% \\
    \midrule
    \multirow{6}[2]{*}{CDs and Vinyl} & Recall@20 & 4.15  & 6.37  & 7.16  & 7.26  & 7.33  & 7.25  & 6.96  & \underline{7.48} & \textbf{8.20} & 9.63\% \\
          & Recall@50 & 5.88  & 8.80  & 9.92  & 10.26  & 10.18  & 10.20  & 10.03  & \underline{10.50} & \textbf{11.41} & 8.67\% \\
          & NDCG@20 & 4.31  & 5.26  & 5.90  & 5.82  & 6.02  & 5.92  & 5.80  & \underline{6.15} & \textbf{6.82} & 10.89\% \\
          & NDCG@50 & 5.04  & 6.17  & 6.85  & 6.35  & 6.78  & 6.54  & 6.64  & \underline{6.92} & \textbf{7.67} & 10.84\% \\
          & HR@20 & 7.21  & 10.57  & \underline{12.35} & 12.23  & 12.28  & 12.26  & 11.87  & 12.16  & \textbf{13.45} & 8.91\% \\
          & HR@50 & 10.87  & 15.60  & \underline{17.36} & 17.05  & 17.34  & 17.24  & 17.10  & 17.15  & \textbf{19.36} & 10.63\% \\
    \bottomrule
    \end{tabular}%
\end{table*}

\section{EXPERIMENTS}
In this section, we perform extensive experiments to evaluate the performance of our proposed DESMIL model.
We need to answer the following Research Questions (RQ):
\begin{itemize}
    \item \textbf{RQ1:} How is the performance of DESMIL under out-of-distribution environments?
    \item \textbf{RQ2:} How is the performance of DESMIL under ordinary in-distribution environments?
    \item \textbf{RQ3:} Is the DESMIL model sensitive to some important hyper-parameters?
    \item \textbf{RQ4:} How does the DESMIL model affect the training process?
    \item \textbf{RQ5:} How is the distribution of sample weights learned in the DESMIL model?
\end{itemize}

\subsection{Experimental Datasets}

We conduct experimental comparison on three public datasets collected from real-world scenarios:

\begin{itemize}
\item \textbf{Book Dataset}. The Book dataset is part of the Amazon Product Data\footnote{\url{http://jmcauley.ucsd.edu/data/amazon/}} \cite{mcauley2015image,he2016ups} in the "book" category. There are 603,668 users, 367,982 items, and 8,898,041 user behaviors in total.

\item \textbf{Movies and TV Dataset}. The Movies and TV dataset is part of the updated version of Amazon Review Data\footnote{\url{https://nijianmo.github.io/amazon/index.html}} \cite{ni2019justifying}. There are 304,763 users, 89,590 items, and 3,506,470 user behaviors in total.

\item \textbf{CDs and Vinyl Dataset}. The CDs and Vinyl dataset is also part of the updated Amazon Review Data. There are 129,237 users, 145,522 items, and 1,682,049 user behaviors in total.
\end{itemize}

Moreover, we need to conduct performance comparison not only in OOD environments for verifying OOD generalization ability, but also in in-distribution environments for verifying ordinary recommendation ability.
Thus, we perform two different data splitting:

(1) The first splitting is \textbf{OOD data splitting}.
As shown in examples in Fig. \ref{fig:examples}, different user groups and popularity tends result in data distribution shift, and affect the recommendation performances.
Considering popularity tends are hard to identify in an offline dataset, we construct OOD data according to different user groups.
In practice, we use the Jaccard similarity\footnote{\url{https://www.learndatasci.com/glossary/jaccard-similarity/}} of items from different users to measure the similarity between users.
We randomly select a user, and then iteratively select the next user with the maximum similarity to the selected users, until $50\%$ users are selected.
We obtain a set of selected users $\mathcal{U}_1$ and a set of remaining users $\mathcal{U}_2$.
In this way, $\mathcal{U}_1$ and $\mathcal{U}_2$ share extremely different distributions, in which $\mathcal{U}_1$ can be viewed as the OOD environment of $\mathcal{U}_2$.
Then, we randomly use $10\%$ in $\mathcal{U}_1$, $10\%$ in $\mathcal{U}_2$ and other non-overlapping $80\%$ in $\mathcal{U}_2$ as the testing set, the validation set and the training set respectively.
That is to say, the number of samples for training, validation and testing confirms to $8:1:1$.

(2) The second splitting is \textbf{random data splitting}. 
We use the same splitting in previous work \cite{cen2020controllable}, in which samples are randomly split into training, validation and testing sets.

\begin{table}[]
\centering
\caption{Comparison Among DESMIL, MGNM+PW and ComiRec+PW under OOD data splitting evaluated by Recall@50 ($\%$). Best performances are indicated by bold font.}
\label{tab:result_PW}
    \begin{tabular}{cccc}
    \toprule
    Approach & Book  & Movies and TV & CDs and Vinyl \\
    \midrule
    ComiRec+PW & 8.63  & 17.16  & 10.25  \\
    MGNM+PW & 8.14  & 17.54  & 10.33  \\
    DESMIL & \textbf{10.89} & \textbf{18.65} & \textbf{11.41} \\
    \bottomrule
    \end{tabular}%
\end{table}

\begin{table*}[t]
\centering
\caption{Results under random data splitting evaluated by different metrics ($\%$). Best performances are indicated by bold font and the strongest baselines are underlined. The improvements indicate the relative increase of DESMIL over the best baselines.}
\label{tab:result_random}
    \begin{tabular}{cccccccccccc}
    \toprule
    Dataset & Metric & GRU4Rec & SASRec & MIND  & ComiRec & SINE  & MGNME  & USR  & CauseRec & DESMIL & Improv. \\
    \midrule
    \multirow{6}[2]{*}{Book} & Recall@20 & 3.47  & 4.76  & 5.10  & 5.92  & 5.46  & \underline{6.18} & 5.75 & 5.24  & \textbf{7.52} & 21.68\% \\
          & Recall@50 & 6.50  & 7.78  & 7.64  & 9.35  & 8.72  & \underline{9.64} & 8.60 & 9.36  & \textbf{11.06} & 14.73\% \\
          & NDCG@20 & 3.55  & 4.84  & \underline{5.09} & 4.17  & 4.83  & 4.88 & 4.93  & 4.66  & \textbf{5.46} & 7.27\% \\
          & NDCG@50 & 4.42  & 5.74  & 5.97  & 5.47  & 6.04  & 6.19 & 5.95  & \underline{6.28} & \textbf{7.24} & 15.29\% \\
          & HR@20 & 7.84  & 8.82  & 10.59  & 11.70  & 11.87  & \underline{12.70} & 11.28 & 12.45  & \textbf{14.86} & 17.01\% \\
          & HR@50 & 12.38  & 13.79  & 15.56  & 18.04  & 18.94  & 20.21 & 17.89  & \underline{20.23} & \textbf{21.53} & 6.43\% \\
    \midrule
    \multirow{6}[2]{*}{Movies and TV} & Recall@20 & 13.20  & 14.43  & 14.87  & 15.46  & 15.16  & \underline{15.50} & 14.76 & 15.30  & \textbf{15.76} & 1.68\% \\
          & Recall@50 & 17.66  & 18.27  & \underline{19.55} & 18.87  & 19.30  & 19.14 & 19.27  & 19.24  & \textbf{20.90} & 6.91\% \\
          & NDCG@20 & 15.07  & 14.49  & \underline{\textbf{15.80}} & 14.73  & 15.57  & 15.44 & 14.90  & 15.10  & 15.54  & - \\
          & NDCG@50 & 16.21  & 16.72  & \underline{17.23} & 16.17  & 16.64  & 16.68 & 16.81  & 16.83  & \textbf{17.36} & 0.75\% \\
          & HR@20 & 22.67  & 23.25  & 25.34  & 25.87  & 25.12  & 25.76 & 24.75  & \underline{25.94} & \textbf{26.42} & 1.85\% \\
          & HR@50 & 29.54  & 30.43  & 32.93  & 33.68  & 33.30  & \underline{33.95} & 32.90 & 33.90  & \textbf{34.80} & 2.50\% \\
    \midrule
    \multirow{6}[2]{*}{CDs and Vinyl} & Recall@20 & 4.39  & 6.92  & 7.55  & 7.96  & 7.69  & \underline{8.03} & 7.35 & 7.77  & \textbf{8.75} & 8.97\% \\
          & Recall@50 & 6.07  & 8.52  & 10.32  & 11.23  & 10.93  & \underline{11.32} & 10.57 & 11.12  & \textbf{12.09} & 6.80\% \\
          & NDCG@20 & 4.81  & 6.44  & \underline{\textbf{7.93}} & 6.84  & 7.28  & 7.46 & 7.12  & 7.51  & 7.86  & - \\
          & NDCG@50 & 5.42  & 7.10  & \underline{8.88} & 8.01  & 8.34  & 8.68 & 8.18  & 8.57  & \textbf{8.91} & 0.34\% \\
          & HR@20 & 8.47  & 12.86  & 14.28  & 14.35  & 14.55  & \underline{14.60} & 14.08 & 14.49  & \textbf{15.73} & 7.74\% \\
          & HR@50 & 11.79  & 16.29  & 19.38  & 20.26  & 20.71  & \underline{20.83} & 19.80 & 20.66  & \textbf{21.89} & 5.09\% \\
    \bottomrule
    \end{tabular}%
\end{table*}

\subsection{Compared Baselines}

We compare our proposed DESMIL model to the following baselines for evaluation:
\begin{itemize}
\item \textbf{GRU4Rec} \cite{hidasi2015session}: a classic sequential recommendation model based on recurrent neural network.
\item \textbf{SASRec} \cite{kang2018self}: a state-of-the-art model that uses self-attention network for the sequential recommendation.
\item \textbf{MIND} \cite{li2019multi}: a classic multi-interest sequential network with dynamic routing for modeling users’ diverse interests in the matching stage.
\item \textbf{ComiRec} \cite{cen2020controllable}: a state-of-the-art sequential network with multi-interest extraction module to generate multiple user interests and aggregation module to obtain top-N items. We use the SA setting of ComiRec which is described as ComiRec-SA in the original paper.
\item \textbf{SINE} \cite{tan2021sparse}: a state-of-the-art multi-interest model that maintains a pool of conceptual prototypes to represent the all set of a user's potential interests, and uses self-attention to decide which prototypes are activated to the user's multiple interests.
\item \textbf{MGNM} \cite{tian2022multi}: a state-of-the-art multi-interest model that combines with graph convolutional networks-based recommenders.
\item \textbf{USR} \cite{wang2022unbiased}: a state-of-the-art debiasing sequential recommendation model with latent confounders in an IPW-based framework, for dealing with exposure bias in user interaction history.
\item \textbf{CauseRec} \cite{zhang2021causerec}: a state-of-the-art sequential network that performs contrastive user representation learning to model the counterfactual data distribution for generalizing to OOD environments.
\end{itemize}
We compare above baselines with DESMIL under both OOD and random data splitting.
Moreover, we additionally consider an IPW-based debiasing method called Permutation Weighting (PW) \cite{arbour2021permutation}.
We perform PW on two representative multi-interest models ComiRec and MGNM, then obtain \textbf{ComiRec+PW} and \textbf{MGNM+PW} for performance comparison with DESMIL in OOD environments.

\subsection{Experimental Settings}
In this subsection, we introduce some details of our experimental settings.
\subsubsection{Parameter Configuration}
The embedding size of items is 64.
According to the best performances of ComiRec, the batch size for the Book dataset is 1024, while for the other two datasets is 128.
The number of negative samples for sampled softmax loss is 10.
All models use early stopping based on the Recall@50 on the validation set.
The de-correlation importance and the number of interests are tuned in the range of $\left\{0.01, 0.1, 1.0, 10.0, 100.0\right\}$ and the range of $\left\{2, 4, 6, 8\right\}$, respectively.
We use the Adam optimizer with learning rate lr = 0.001 for optimization.\\
\subsubsection{Evaluation Metrics}
We use the top-$p$ Recall, Normalized Discounted Cumulative Gain (NDCG), and Hit Rate (HR) to evaluate performances of all the compared models. We select $p=20, 50$ in our experiments. The three metrics measure the model performances with different criteria.
Recall@$p$ is defined as the fraction of relevant items found in the top $p$ recommended items. NDCG@$p$ further considers the normalization of gains and the ranking of correctly recommended items, where items with higher relevance affect the final score more. HR@$p$ is defined as the proportion of top $p$ recommended items found in the testing set. 

\subsection{Results under OOD Data Splitting (RQ1)}

In Tab. \ref{tab:result_OOD}, we illustrate the experimental comparison under the OOD data splitting.
Among the compared baselines, CauseRec performs the best, achieves best performances on 12 out of 18 target metrics.
This shows that, to a certain extent, training with counterfactual samples can improve the model's OOD generalization ability.
USR performs better than single-interest models, and is competitive with multi-interest models.
Considering the recommendation backbone in USR is a simple single-interest GRU, this shows the effectiveness of IPW-based approaches.
Among multi-interest models, simple models, i.e., MIND and ComiRec, seem to have relatively better performances.
This indicates that, models with simpler structures tend to perform better in different OOD environments.
Moreover, our proposed DESMIL model clearly outperforms all the compared baselines by large margins.

Meanwhile, we conduct comparison among DESMIL, MGNM+PW and ComiRec+PW under the OOD data splitting in Tab. \ref{tab:result_PW}.
We can observe that, among the compared baselines, ComiRec+PW performs better on the Book dataset, and MGNM+PW performs better on the CDs and Vinyl dataset and the Movies and TV dataset.
Compared with results in Tab. \ref{tab:result_OOD}, MGNM+PW and ComiRec+PW perform better than MGNM and ComiRec respectively, and are competitive with CauseRec.
This further shows the effectiveness of IPW-based approaches \cite{schnabel2016recommendations,wang2019doubly,arbour2021permutation} for OOD generalization.
And DESMIL still significantly outperforms above two PW-augmented approaches.
These results and observations strongly demonstrate the effectiveness and stability of DESMIL in OOD environments.

\begin{figure}
    \centering
    \includegraphics[width=0.95\linewidth]{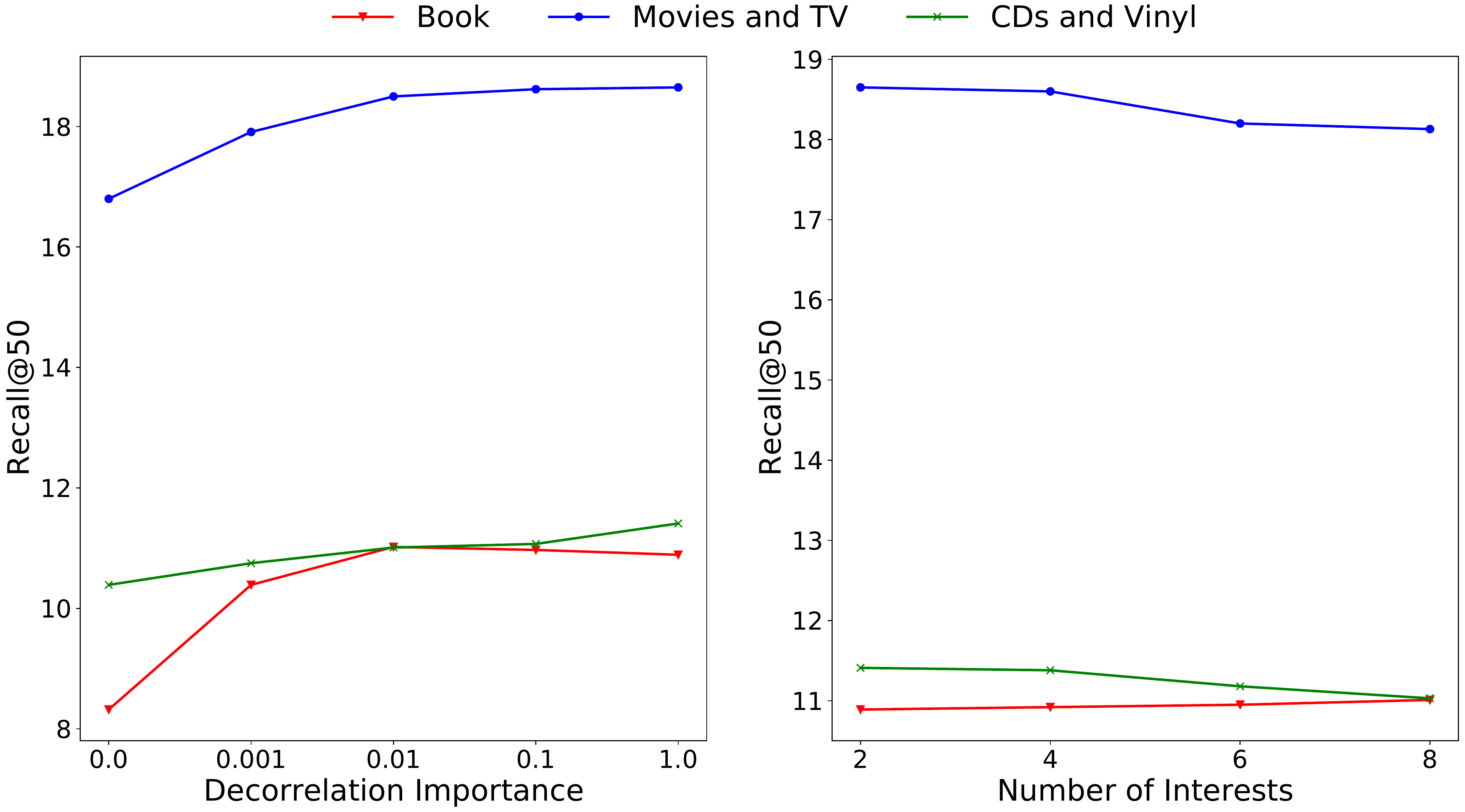}
    \caption{Hyper-parameter study of de-correlation importance coefficient and number of interests under OOD data splitting measured by Recall@50 ($\%$).}
    \label{fig:params_ood}
\end{figure}

\begin{figure}
    \centering
    \includegraphics[width=0.95\linewidth]{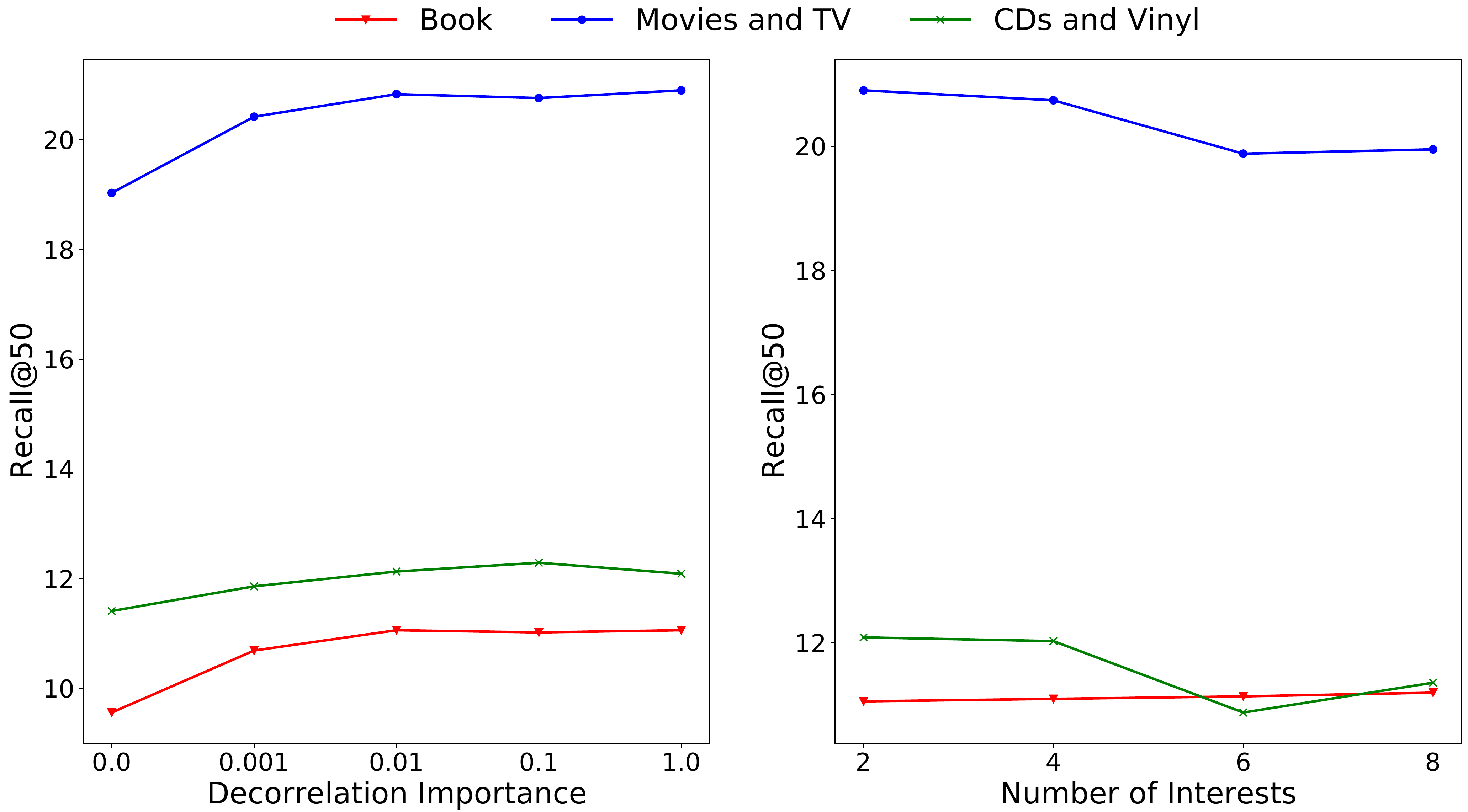}
    \caption{Hyper-parameter study of de-correlation importance coefficient and number of interests under random data splitting measured by Recall@50 ($\%$).}
    \label{fig:params_rnd}
\end{figure}

\subsection{Results under Random Data Splitting (RQ2)}

The experimental comparison under the random data splitting is shown in Tab. \ref{tab:result_random}.
We can observe that, results in Tab. \ref{tab:result_OOD} are commonly lower than those in Tab. \ref{tab:result_random}, which shows distribution shift leads to more difficult tasks.
Overall speaking, attention-based SASRec performs better than RNN-based GRU4Rec, and multi-interest models have better performances than both of them.
With the help of graph convolutional networks, MGNM is the best one among the compared multi-interest models.
Meanwhile, our proposed DESMIL model constantly outperforms all the compared baselines, except evaluated by NDCG@20 on the Movies and TV dataset and the CDs and Vinyl dataset.
On the Book dataset, and evaluated by Recall and HR, the improvements of DESMIL are significant.
In real applications, Recall is often considered the most important metric as it can best reflect the model performance facing an enormous candidate set of items and almost equally important but limited exposure positions.
These results further demonstrate the effectiveness of our proposed DESMIL model.
Moreover, comparing with results in Tab. \ref{tab:result_OOD} and Tab. \ref{tab:result_random}, improvements achieved by DESMIL under OOD settings are much more significant than those under random settings.
This indicates that, DESMIL is a multi-interest model suitable to OOD generalization, but still has great performances under regular data splitting.

\begin{figure}
    \centering
    \includegraphics[width=0.95\linewidth]{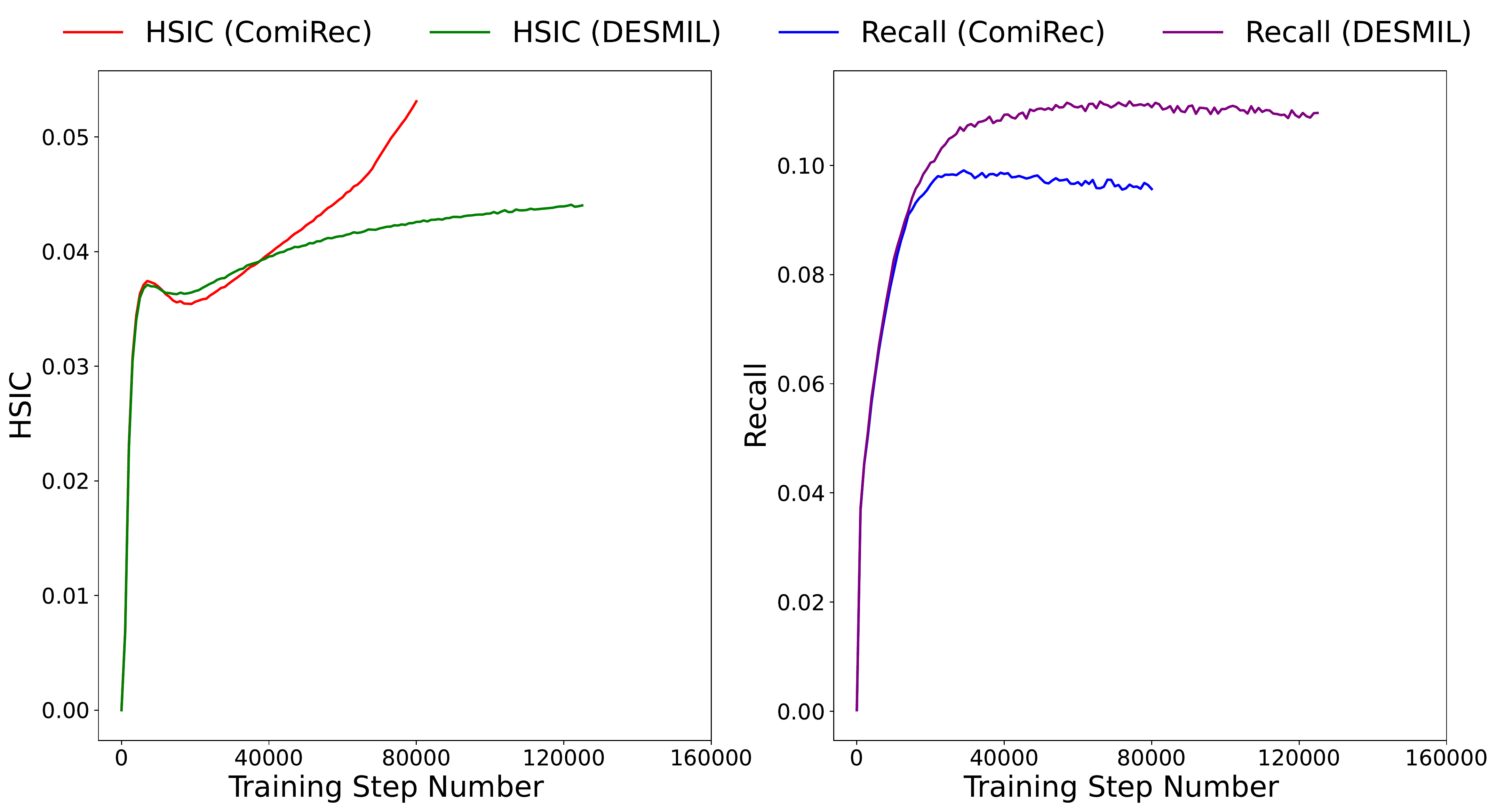}
    \caption{The curves of HSIC on the training set and Recall@50 on the validation set when training ComiRec and DESMIL on Book. Compared with ComiRec, DESMIL shows slighter correlations among multiple interests measured by HSIC, and better performance measured by Recall@50.}
    \label{fig:book_vis}
\end{figure}

\begin{figure}
    \centering
    \includegraphics[width=0.8\linewidth]{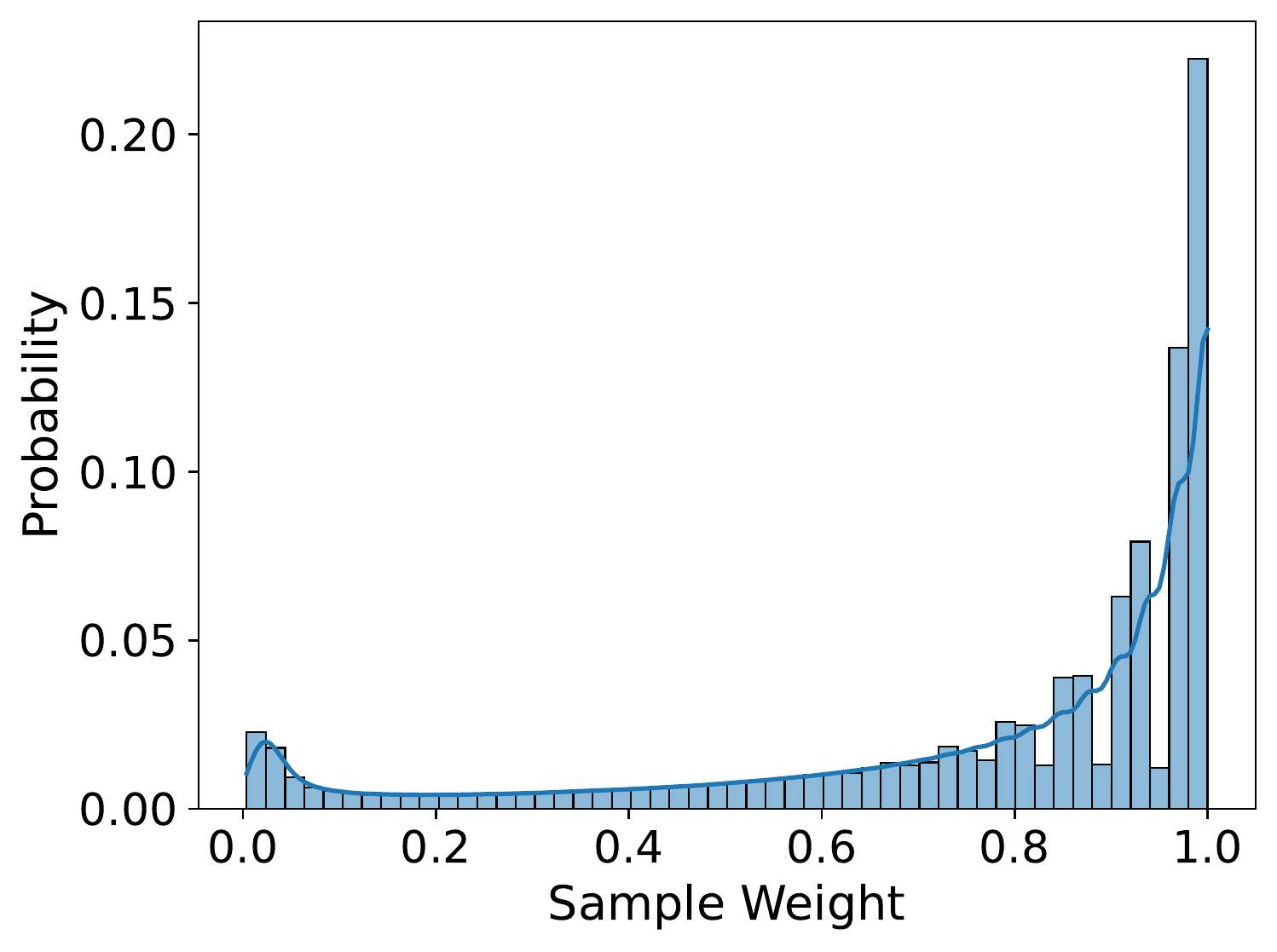}
    \caption{The distribution of sample weights in the DESMIL model trained on the Book dataset.}
    \label{fig:book_sample_weight}
\end{figure}

\subsection{Hyper-parameter Sensitivity (RQ3)}

To investigate the stability of our proposed model to the hyper-parameters, we conduct hyper-parameter sensitivity study.
In Fig. \ref{fig:params_ood} and Fig. \ref{fig:params_rnd}, we illustrate the sensitivity of de-correlation importance coefficient $\lambda$ and number of interests $c$, under the OOD data splitting and the random data splitting respectively.
Results in the figures are evaluated by the Recall@50 metric.
We can observe that, $c$ does not affect the performances of DESMIL very much.
Moreover, when $\lambda = 0.0$, performances of DESMIL drop significantly.
And $\lambda = 0.0$ indicates DESMIL without interest de-correlation, which can also be viewed as the ablation study.
This shows the importance of interest de-correlation in DESMIL.
Meanwhile, when $\lambda \in \left[ 0.01, 0.1, 1.0 \right]$, the performances of DESMIL stay relatively stable, which shows that we do not have too much burden for hyper-parameter tuning in practice.
In our other experiments, we simply set $\lambda=1.0$ and $c=2$.

\subsection{Visualization (RQ4 and RQ5)}

In Figure \ref{fig:book_vis}, we visualize the change of HSIC on the training set, and Recall on the validation set when training DESMIL and ComiRec on the Book dataset. 
Both DESMIL and ComiRec use early stopping and their training terminates at different steps, which results in the different lengths of curves shown in the figure.
To be noted, DESMIL performs optimization of HSIC by sample re-weighting in the training phase, while the calculation of HSIC, which is shown in the figure, is not weighted.
Different from DESMIL, ComiRec does not control the optimization of HSIC, i.e., correlations among interests, during training.
During the first 10000 steps, the HSIC and Recall of both models quickly increase.
Then, the HSIC of ComiRec continues to increase rapidly, while the HSIC of DESMIL grows relatively smoothly.
This makes it possible for DESMIL to update more steps and obtain better performances.
In a word, via minimizing the weighted correlation estimation loss based on HSIC in Eq. (\ref{eq:correlation_loss}), we can break the trade-off relation between the correlations among interests and the model performances introduced in Fig. \ref{fig:book_pilot}, and alleviate the dependencies between stable interests and noisy interests which may mislead the model to learn spurious correlations.

Moreover, in Figure \ref{fig:book_sample_weight}, we illustrate the probability distribution of sample weights leaned in DESILL on the Book dataset, in the form of histograms.
Sample weights in the figure are mostly in the values from $0.8$ to $1.0$, with some located near the value of $0.0$.
The values near $1.0$ indicate slight changes in sample weights, while the values near $0.0$ indicate sharp changes of sample weights in the main objective loss function.
This shows that, most samples in the Book dataset do not require specific de-correlation operations, while a small part of samples are indeed marginalized.

\section{Conclusion}
In this paper, we investigate the OOD generalization problem in multi-interest models, for accurate and stable sequential recommendation.
To achieve this, we propose a novel multi-interest recommendation model called DESMIL.
DESMIL uses an attentive module to extract multiple interests, and selects the most important one for conducting final predictions.
Then, a weighted correlation estimation loss is incorporated, to alleviate the correlations among different extracted interests in the training set.
The DESMIL model can learn stable representations in sequential recommendation, and make stable and accurate predictions generalized to OOD environments.
Extensive experimental results under both OOD and random settings strongly demonstrate that our proposed DESMIL model is a promising sequential recommendation model.



\ifCLASSOPTIONcaptionsoff
  \newpage
\fi

\balance
\bibliographystyle{IEEEtran}
\bibliography{IEEEtran.bib}

\end{document}